# Straddling Two Platforms: From Twitter to Mastodon, an Analysis of the Evolution of an Unfinished Social Media Migration

Simón Peña-Fernández[1*], Ainara Larrondo-Ureta[2] and Jordi Morales-i-Gras[3]

1. University of the Basque Country (UPV/EHU) (ROR: 000xsnr85)
   simon.pena@ehu.eus / https://orcid.org/0000-0003-2080-3241
2. University of the Basque Country (UPV/EHU) (ROR: 000xsnr85)
   ainara.larrondo@ehu.eus / https://orcid.org/0000-0003-3303-4330
3. University of the Basque Country (UPV/EHU) (ROR: 000xsnr85)
   jordi.morales@ehu.eus / https://orcid.org/0000-0003-4173-3609

\* Corresponding author

**Abstract**: Social media have been fundamental in the daily lives of millions of people, but they have raised concerns about content moderation policies, the management of personal data, and their commercial exploitation. The acquisition of Twitter (now X) by Elon Musk in 2022 generated concerns among Twitter users regarding changes in the platform's direction, prompting a migration campaign by some user groups to the federated network Mastodon. This study reviews the onboarding of users to this decentralised platform between 2016 and 2022 and analyses the migration of 19,000 users who identified themselves as supporters of the platform switch. The results show that the migration campaign was a reactive response to Elon Musk's acquisition of Twitter and was led by a group of highly active academics, scientists, and journalists. However, a complete transition was not realised, as users preferred to straddle their presence on both platforms. Mastodon's decentralisation made it difficult to exactly replicate Twitter's communities, resulting in a partial loss of these users' social capital and greater fragmentation of these user communities, which highlights the intrinsic differences between both platforms.

## 1. Introduction

Since their emergence in the mid-2000s, social media have become an integral part of daily life for millions of people (Pew Research Center 2024). Their multidimensional nature has allowed them to quickly adapt to various environments (personal and social, public and private, work and leisure, etc.), leading to their rapid spread.

Social media are meeting points with large audiences that enable the quick publication of content and its distribution through networks created by the users themselves, who can, in turn, interact with content created by others by recommending, redistributing, or commenting on it (Boyd and Ellison 2007; Kupferschmidt 2022).

In their development, the most successful platforms have been centralised networks, meaning those that are owned by a single company (Al-Khateeb 2022), where all tasks—identity, authentication, data storage, addressing, governance, content moderation, etc.—are managed by a single provider (Abbing et al. 2023).

However, at the same time that the general public has embraced the use of mainstream social media, their data management policies have also raised concerns among users. Scandals like the Cambridge Analytica case led many users to express concern about the privacy of their data and how it is used (Guidi et al. 2020).

This exploitation of data for commercial purposes, passively accepted by users, has maximised targeted ads and content personalisation, also contributing to the creation of information disorders such as filter bubbles or echo chambers (La Cava et al. 2022a). Many users also began to perceive that social media had moved beyond their original purpose as interconnection platforms and were dedicated to commercially exploiting their connections. All of this was surrounded by a heated debate about toxicity in social media (Boyle et al. 2021), with highly polarised positions between those who feel that not enough is being done to protect users from extreme or undesirable content and, on the other hand, those who believe these policies limit freedom of expression (Anderson 2017).

For users seeking alternatives to Twitter, Mastodon represents an appealing option. Launched in October 2016, this microblogging platform is a federated social network that fosters diversity in users, small but interconnected communities, and user ownership of their data (Zignani et al. 2018; Shaw 2020). This type of server federation, commonly referred to as the Fediverse, is based in the interconnection of independent servers (instances) that share common protocols, which allows for the exchange of information across different instances regardless of the software being used on each (Raman et al. 2019; La Cava et al. 2021; Lázaro-Rodríguez 2024).

Mastodon allows users to write messages (toots) up to 500 characters long and interact with them in a very similar way to Twitter, by reposting, sharing, and more. As other decentralised platforms, Mastodon is built on web standards and open-source tools (La Cava et al. 2022b). In a way, they aim to reclaim the original spirit that drove the internet as a decentralised network (Boyle et al. 2021; Brembs et al. 2023; Orihuela 2023).

From the perspective of privacy protection, Mastodon offers its users much more control (Zulli et al. 2020; Wang 2021). In contrast to other platforms, data are only stored on the server where the account resides. Additionally, due to its federated nature, switching instances is straightforward for those who disagree with the policies of the server they are subscribed to (Shaw 2020). This ensures greater transparency and eliminates the use of recommendation or monetisation systems that would otherwise influence users in a conditioned way. Like other decentralised initiatives, Mastodon is non-commercial and ad-free, meaning it is not driven by profit but rather supported through voluntary work, donations, and sponsorships (Laser et al. 2023).

However, as a social media, Mastodon also has some limitations (Braun 2023). On one hand, it is difficult to have a global view of the network's content, as users can only read public messages shared by other users within their own instance and the content posted by users they follow from other instances, but they cannot access the entire set of messages published (Shaw 2020). On the other hand, this very federated nature can make the eradication of questionable content a difficult task, as it depends on the willingness of the server administrators hosting it (Abbing et al. 2023).

In the same vein, although the diversity between servers is high, the federated nature also seems to produce, in some cases, a certain insularity, where relationships occur between users of the same instance or from instances in the same country (Zignani et al. 2019; Erz 2022). This can lead to a fractured perception of reality, where users interact in silos with people who share their tastes and opinions (Shaw 2020). Despite its federated nature, users coming from commercial social media tend to instinctively sign up for the largest server (mastodon.social). This effect results in 52% of users being registered on only 10% of the servers (Raman et al. 2019).

With a structure and objectives so different from those of commercial platforms, Mastodon had existed since its creation in 2016 as a peripheral and residual alternative to Twitter. However, its opportunity as a real alternative to Twitter came with the rumors, and the subsequent confirmation, of Elon Musk's acquisition of Twitter. It was then that some users launched a campaign advocating for migration, using hashtags like #GoodbyeTwitter and #TwitterMigration, or including their Mastodon account details in their profile descriptions.

Among their motivations was the accumulated deterioration of their relationship with Twitter, which led long-established communities to become disillusioned and seek alternatives (Swogger 2023). A more direct influence came from Musk's own statements regarding his new moderation policies and the potential impact of cuts and layoffs on the global conversation as well as the introduction of paid account verification (Kupferschmidt 2022), questions for which Mastodon seemed a natural alternative (Lee and Wang 2023; Sabo and Gesthuizen 2024).

In this context, the main objective of this study is to analyse the migration of users from Twitter to Mastodon and, in particular, the impact of their networks.

Based on this general objective, the following specific objectives are proposed:
RQ1. What has been the temporal evolution in the onboarding of users to Mastodon, and what have been its main milestones?
RQ2. Is there any correlation between the change in ownership of Twitter and the migration to Mastodon?
RQ3. What are the characteristics of the communities of users with accounts on both social media?
RQ4. What is the impact of migration on users' networks? Can we speak of a migration of virtual communities?

## 2. Theoretical Framework

The study of migration on social media has a broad academic trajectory. From the earliest investigations into users' movements between platforms, the loss or reconfiguration of digital social capital, to the effects of technological innovations and moderation policies, the literature has addressed these phenomena as deeply social processes, going beyond mere shifts in technical infrastructure.

In their adaptation of the push–pull–mooring (PPM) theory, Bansal et al. (2005) transferred this migratory framework, originally developed in human geography, to the context of consumer behaviour, specifically for analysing service provider switching in digital environments. The push–

pull–mooring model explains switching behaviour through three types of factors: push factors, which drive users away from their current provider; pull factors, which attract them to alternatives; and moorings, which either facilitate or inhibit change. According to this model, even when users are dissatisfied or perceive attractive alternatives, strong moorings—such as switching costs or established habits—can play a decisive role in maintaining their loyalty.

The PPM model has been widely adopted to explain migrations within the digital environment (Handarkho and Harjoseputro 2020; Lin et al. 2021; Xia et al. 2023), and in all of these cases, mooring effects play a crucial role. For example, inertia can be one of the main factors determining whether users ultimately decide to stay, even when push and pull forces point towards change. It serves as a stabilising mechanism, amplifying the influence of emotional attachment, perceived switching costs, and habitual behaviours, thereby reducing the impact of dissatisfaction or attractive alternatives (Sun et al. 2017).

The PPM model has also been applied specifically to the case of Twitter. Jeong et al. (2024a) highlight that factors such as dissatisfaction with governance (push), attraction to decentralised structures (pull), and users' prior social capital and habits (mooring) interact to shape digital migration trajectories.

In line with the strong influence of mooring factors in the PPM theory, Solomonik and Heuer (2025) point out that migration across social platforms is driven more by social, emotional, and contextual factors than by technical functions. Peer popularity, local perceptions of safety, and the alignment of platforms with life stages all influence decisions. According to these authors, such migrations reflect personal trajectories rather than purely rational choices.

Standing out in these migration processes are users who hold a prominent leadership position. As Kumar et al. (2011) note, users most likely to migrate between social networks are those who display high levels of activity, maintain extensive social networks (i.e., many friends or followers), and enjoy high external visibility. These users, referred to as high net-worth individuals (HNWI), are particularly influential within the platform ecosystem and often set migration trends. Moreover, when these users become inactive, they tend to precede broader waves of abandonment, making their behaviour a useful indicator for predicting shifts in a platform's popularity.

The diffusion of innovations theory (Rogers 2003), for its part, explains digital migration not only through the influence of early adopters, but also through perceived relative advantage, compatibility with prior usage, technical complexity, trialability, observability of benefits, and social pressure—all of which shape the adoption of new platforms. These factors mean that innovation does not occur uniformly but rather gradually and unevenly, shaped by communicative, social, and cultural influences. The process depends not only on the characteristics of the innovation itself but also on the social context in which it circulates.

## 3. Materials and Methods

To conduct this analysis, a preliminary list of over 60,000 Twitter users was compiled, all of whom stated on their Twitter profiles that they also had a Mastodon account, either in their username or profile description. This identification was carried out by searching the Twitter academic API at the end of 2022 for users on Twitter who shared their Mastodon profile. The identification process was conducted using a search on the Twitter Academic API 1.1 in November 2022. After analysing the selected data, a total of 19,919 users with accounts on both platforms were identified.

First, the data regarding the migration process from one platform to another (RQ1) were analysed, specifically, the registration date of the selected users on Mastodon during the period between 1 October 2016 and 2 November 2022.

Secondly, to try to establish the correlation between the change in Twitter's ownership and migration to Mastodon (RQ2), the evolution of registration data on this platform was compared with the evolution of tweets containing the words "Musk" and "Twitter" that had achieved 1000 retweets during the same analysis period.

For the analysis of the communities existing on both social media (RQ3), the following relationships of the users analysed were obtained from Twitter using API 1.1. In contrast, on Mastodon, although it is technically possible to retrieve the data through the official API (Schoch and Chan 2023; Bin Zia et al. 2024), the relational data were collected using web scraping techniques, parsing public HTML from user profiles with standard Python libraries like Requests (v. 2.29.0) and BeautifulSoup (v. 4.12.2). This approach was necessary due to difficulties in consistently obtaining comprehensive follower graphs across the federated network for our specific user set via the API at the time of collection. Consequently, this data collection focuses on publicly available relational information, excluding connections involving private instances or protected accounts. These collected relational data were used to construct follower graphs for both platforms. We then employed Social Network Analysis (SNA) using the Python library NetworkX (v. 3.1) and Gephi (v. 0.10.1) to calculate key network metrics (including density, clustering coefficient, modularity, centralisation, average degree, and number of communities, as detailed in Table 1) and to detect communities using the Louvain algorithm (resolution = 1). Categorical assortativity was also calculated to assess community preservation.

Table 1. Characteristics of the communities on Twitter and Mastodon.

|  | **Twitter Followers Network** |  | **Mastodon Followers Network** |  |
|---|---|---|---|---|
| Before October 2022 4.583 nodes | Density | 0.002 | Density | 0.001 |
|  | Clustering coefficient | 0.14 | Clustering coefficient | 0.087 |
|  | Modularity | 0.524 | Modularity | 0.509 |
|  | No. of communities | 283 | No. of communities | 732 |
|  | Indegree centralisation | 0.135 | Indegree centralisation | 0.187 |
|  | Outdegree centralisation | 0.022 | Outdegree centralisation | 0.046 |
|  | Average degree | 8.537 | Average degree | 5.436 |
|  |  |  | Twitter Communities' assortativity | 0.667 |
| After October 2022 15.336 nodes | Density | 0.001 | Density | 0.001 |
|  | Clustering coefficient | 0.13 | Clustering coefficient | 0.091 |
|  | Modularity | 0.542 | Modularity | 0.528 |
|  | No. of communities | 278 | No. of communities | 1141 |
|  | Indegree centralisation | 0.137 | Indegree centralisation | 0.149 |
|  | Outdegree centralisation | 0.013 | Outdegree centralisation | 0.021 |
|  | Average degree | 15.182 | Average degree | 9.193 |
|  |  |  | Twitter Communities' assortativity | 0.584 |

Source: Author's own work.

Finally, we measured whether the move to Mastodon resulted in abandoning Twitter in order to determine whether we can speak of migration or the simultaneous use of both platforms (RQ4). To do this, tweets from a sample of roughly 2000 Twitter users were captured—guaranteeing between 200 and 300 users per cluster of inter-est—over a period of ± 1 month from their registration date on Mastodon (60 days in total) to assess whether a change in their activity level could be observed.

## 4. Results

*4.1. Evolution of Mastodon Sign-Ups and the "Elon Musk Effect"*

The analysis of the Mastodon registrations, first and foremost, shows that this is a shift that occurs at specific moments, driven by bursts and in batches, rather than gradually and steadily over time. The analysis of the evolution of sign-ups on the platform reveals five clearly differentiated peaks (Figure 1). Although Mastodon was launched in October 2016, the first significant peak in user registrations did not occur until April 2017. The second surge in sign-ups took place in November of that same year, when some prominent figures in social media announced their move to this platform. Among them were artists, writers, and tech personalities and entrepreneurs such as Chuck Wendig, John Scalzi, Melanie Gillman, or John O'Nolan. A significant third peak occurred in August 2018.

Figure 1. Evolution of Mastodon registrations.

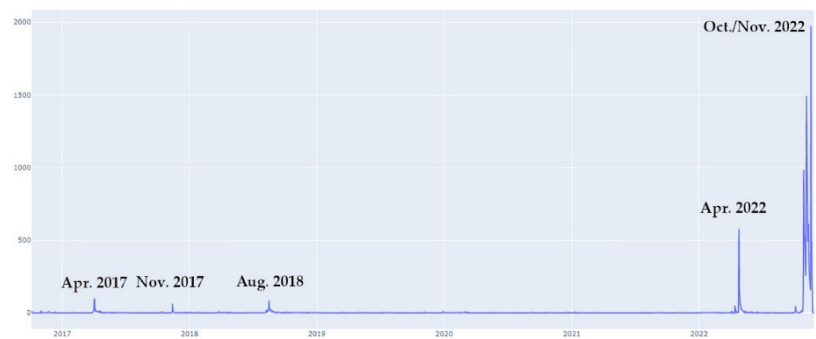

Source: Author's own work.

In any case, during its early years of existence, access to Mastodon could be considered marginal compared to the number of active users on commercial social media. Although initial surges of users were already observed as a result of the first campaigns to leave Twitter, the influx of users to Mastodon remained modest over the next four years.

However, coinciding with the announcement on 25 April 2022 of Elon Musk's purchase of Twitter, the flow of new registrations on Mastodon surged with the largest influx of users to the platform since its creation, and six months later, after a heated legal controversy over the terms of the purchase, Elon Musk officially acquired Twitter on 28 October 2022 for USD 44 billion. After that, sign-ups to Mastodon skyrocketed.

In particular, in the last two cases, a clear correlation can be observed between the change in Twitter's ownership and the beginning of the migration to Mastodon. Regardless of the possible reasons speculated to explain this shift—whether it is concerns over changes in the platform's policies, fear of losing its spirit, or rejection of the new owner—the connection between these two events is clear. During the period from 25 April to 1 May 2022, 7.27% of the users we tracked were registered, and the figure rises to 75.64% in the period from 27 October to 22 November 2022.

To reinforce this analysis and visualise the temporal relationship, Figure 2 plots both the daily Mastodon registrations from the sampled Twitter users (left Y-axis) and the daily count of high-impact tweets related to 'Musk' and 'Twitter' (right Y-axis) over the analysis period. The figure illustrates the sharp increases in both series, particularly during the key periods in 2022. The statistical analysis of both variables shows a moderate–high positive correlation ($R = 0.737$), meaning they increase together. Furthermore, the p-value $< 0.0001$ indicates that the result is highly significant, meaning it is very unlikely that this correlation occurred by chance.

Figure 2. Correlation between Mastodon registrations and activity on Twitter.

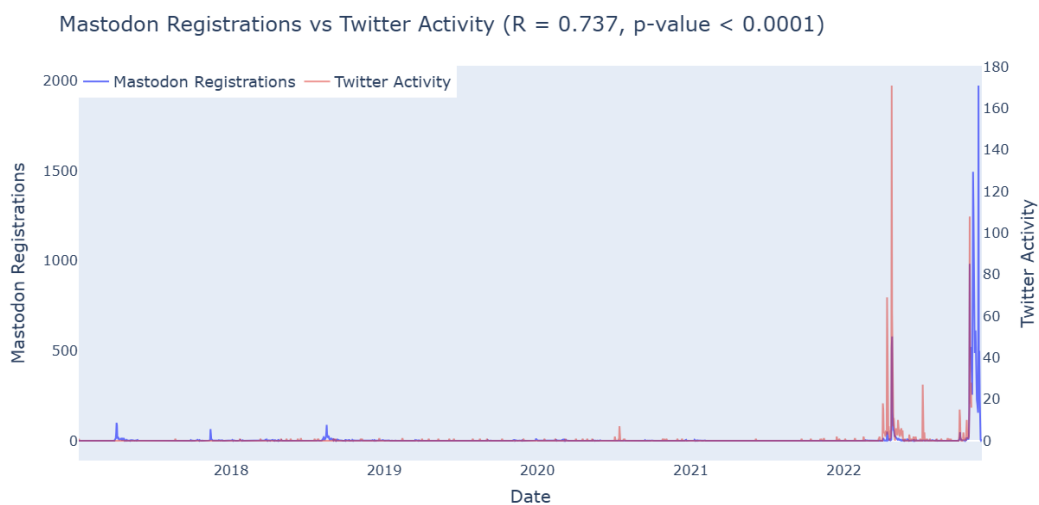

Source: Author's own work.

While this strong positive correlation indicates a clear temporal association between heightened discussion about Musk's acquisition on Twitter and the surge in Mastodon sign-ups among these users, establishing direct causality would require controlling for external factors, such as broader media coverage or pre-existing user dissatisfaction, which is beyond the scope of this specific analysis. Nonetheless, the data strongly suggest that the migration wave had a significant reactive component linked to the change in Twitter's ownership. When the debate over Elon Musk's purchase of Twitter intensified within the social media itself, there was an explosion of user registrations on Mastodon.

*4.2. Interaction and Cohesion of the Communities and Relational Capital*

Regardless of the reasons behind it, the case of Twitter and Mastodon is of particular interest, as it allows us to characterise how user and virtual community migrations occur on social media.

The shift from one social media to another or the adoption of a new platform does not simply involve a personal user registration on the new platform. Social media are the tools through which individuals create and participate in virtual communities, a relational capital they also wish to transfer to the new platform, particularly the most active ones.

Therefore, Twitter users who expressed their intention to migrate tried to bring their relational capital to Mastodon as well. In other words, they sought to rebuild on Mastodon the virtual community they had previously established on Twitter.

We created two versions of the Twitter and Mastodon follower networks, distinguishing between users who registered on Mastodon before or after October 2022, resulting in four networks in total

(Table 1). This allowed us to compare the structure of the Twitter follower network of each group with its corresponding Mastodon network to assess how the migration affected users' connectivity and community structure. Our analysis indicates that the migration from Twitter to Mastodon has been associated, both before and after October 2022, with several structural changes in the network, affecting cohesion, fragmentation, and centralisation.

The density of a network measures how interconnected its nodes are by comparing the number of existing connections to the maximum possible number. A higher density indicates a more tightly connected network, while a lower density suggests a more dispersed structure. Among users who registered on Mastodon before October 2022, the Twitter network had a density of 0.002, while the Mastodon network showed a lower value of 0.001. For those who registered after October 2022, the density in Twitter was also 0.001, remaining the same in Mastodon. These values indicate that despite the migration, the overall level of connectivity among users did not greatly decrease in Mastodon, suggesting that while users re-established their connections, the overall structure remained very similar to Twitter's.

Despite this initial insight, several metrics suggest that Mastodon re-elaborated networks are less cohesive and more fragmented than originally in Twitter. For instance, the clustering coefficient measures the likelihood that a user's connections are also connected to each other, forming closed triangles. A higher clustering coefficient suggests that users belong to tightly knit communities. Among users who joined Mastodon before October 2022, the Twitter network had a clustering coefficient of 0.14, whereas the corresponding Mastodon network showed a lower value of 0.087. For those who migrated after October 2022, the Twitter network's clustering coefficient was slightly lower at 0.13, while the Mastodon network had a value of 0.091. The consistently lower clustering coefficient in Mastodon suggests that regardless of when users migrated, their networks became less cohesive compared to Twitter, with fewer strongly interconnected groups.

In the same vein, the average degree, which represents the mean number of connections per user, illustrates differences in connectivity between the platforms. Among early adopters (i.e., before October, 2022), the average degree in Twitter was 8.537, while in Mastodon, it was lower at 5.436. Among later adopters, the average degree also decreased from 15.182 in Twitter to 9.193 in Mastodon. These values indicate that users who migrated later tended to belong to more cohesive communities, but their Mastodon networks remained less connected compared to their corresponding Twitter networks in any case.

The number of communities detected with the Louvain algorithm with resolution = 1—which have yielded very similar values of Modularity in the four networks, with figures ranging from 0.509 to 0.542—provides additional insights into network fragmentation. Among early adopters, Twitter had 283 communities, while Mastodon had more than twice that number, with 732. Among later adopters, Twitter showed a similar number of communities, with 278, whereas in Mastodon, the number of communities increased sharply to 1141. This significant rise suggests that late adopters on Mastodon formed even more fragmented communities than early adopters, but the migration from one medium to the other implied in both cases a decrease in network cohesion and an increase in fragmentation.

On the other hand, indegree and outdegree centralisation metrics measure the extent to which certain nodes accumulate more connections than others, indicating hierarchical structures within the network. Higher centralisation values suggest that a few users hold disproportionate influence over the network. Among early adopters, indegree centralisation in Twitter was 0.135, while in

Mastodon, it was slightly higher at 0.187. Among later adopters, this value increased marginally, reaching 0.137 in Twitter and 0.149 in Mastodon. Similarly, outdegree centralisation among early adopters was 0.022 in Twitter and 0.046 in Mastodon, while among later adopters, it was 0.013 in Twitter and 0.021 in Mastodon. The consistently higher centralisation in Mastodon suggests that regardless of when users migrated, the network became slightly more hierarchical, with certain accounts accumulating a larger share of the connections.

An additional measure, Twitter communities' assortativity in Mastodon, evaluates whether users who belonged to the same community on Twitter tend to remain connected in Mastodon. Among early adopters, this value was 0.667, while for later adopters, it decreased slightly to 0.584. These values suggest that Twitter communities were largely preserved in Mastodon, although there was a slight decline in community structure among later adopters.

All these findings suggest that although the migration allowed users to retain an important part of their relational capital, as suggested by metrics such as density or categorical assortativity, the migration from Twitter to Mastodon also led to important structural changes in the network. The decrease in density and clustering coefficients, combined with the increase in the number of detected communities, indicates that the Mastodon network became less cohesive and more fragmented compared to Twitter. At the same time, the increase in indegree centralisation suggests that Mastodon's network is not only more decentralised overall but also exhibits some degree of hierarchy, where certain accounts hold more influence.

In summary, the transfer (or recreation) of Twitter's virtual communities to Mastodon has resulted in a significant loss of relational capital. It should be noted that this study was conducted among the most active and motivated users and that the process took place in batches—meaning it occurred on a large scale within a short period of time. Twitter users who announce their presence on Mastodon transfer part of their relational capital to the new social media, although the migration reduces the size of the community and its cohesion while increasing the total number of virtual communities on the new platform.

*4.3. Typology of the Main Migrated Communities*

From the analysis of the user clusters common to both platforms (Figure 3), it is also possible to determine how the pre-existing communities from Twitter have been (re)created upon migrating to Mastodon. This process is detailed in Table 2 and visualised in the Sankey diagram presented in Figure 4. Figure 4 illustrates these community transitions using a Sankey diagram. The left side represents the main user communities identified on Twitter (labelled C.X—Twitter), and the right side shows the communities identified on Mastodon (labelled C.Y—Mastodon). The width of the flows connecting communities indicates the proportion of users moving from a specific Twitter community to a specific Mastodon community. This visualisation highlights both the tendency for communities to replicate (e.g., 'Digital activists and journalists USA') and instances of regrouping or fragmentation into new or multiple Mastodon communities (e.g., users from 'Scientists and professors USA' distributing across several Mastodon groups). As the data show, the vast majority of the communities detected in Mastodon using the Louvain algorithm replicate in a percentage of over 70% some community that already existed on Twitter and that we have proceeded to label with the same name; this is the case of the communities Independent journalists USA, Digital activists and journalists USA, University professors GER, Scientists and professors UK, Scientists and professors USA, Scientists and professors FRA, Scientists and professors BEL and NL, and Scientists and professors ESP. Other Mastodon communities, such as 6 (Scientists and professors USA and EUR) or 9 (Astronomers and physicists USA), regroup users who were in several communities, with Twitter community 8 (Scientists and professors USA) being the main one in both cases. On the other hand, Mastodon community 13 (Sci. and prof. EUR, CAN, and AUS) comes mainly from Twitter community 3 (Environmental sci. and prof. USA) but also regroups users from other communities.

Figure 3. Comparison of clusters on Twitter and Mastodon.

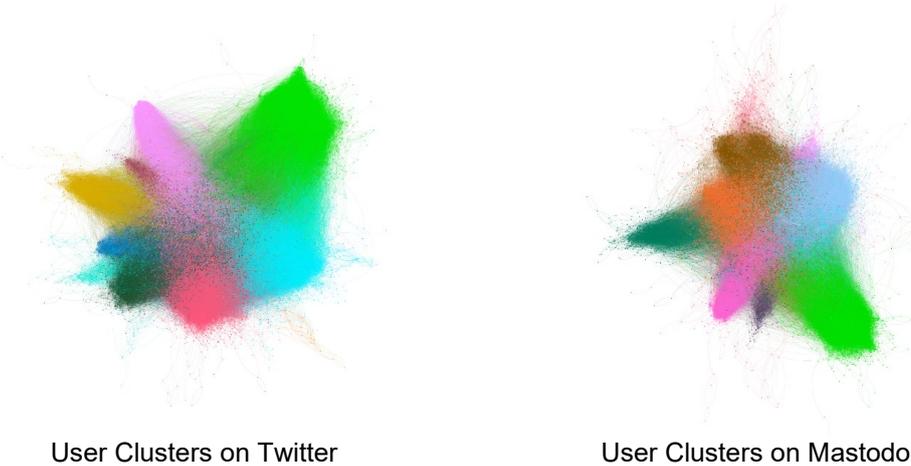

User Clusters on Twitter    User Clusters on Mastodon

Source: Author's own work.
Note: Colors represent different user clusters in the network.

Figure 4. The Sankey diagram with the user flows from Twitter to Mastodon communities.

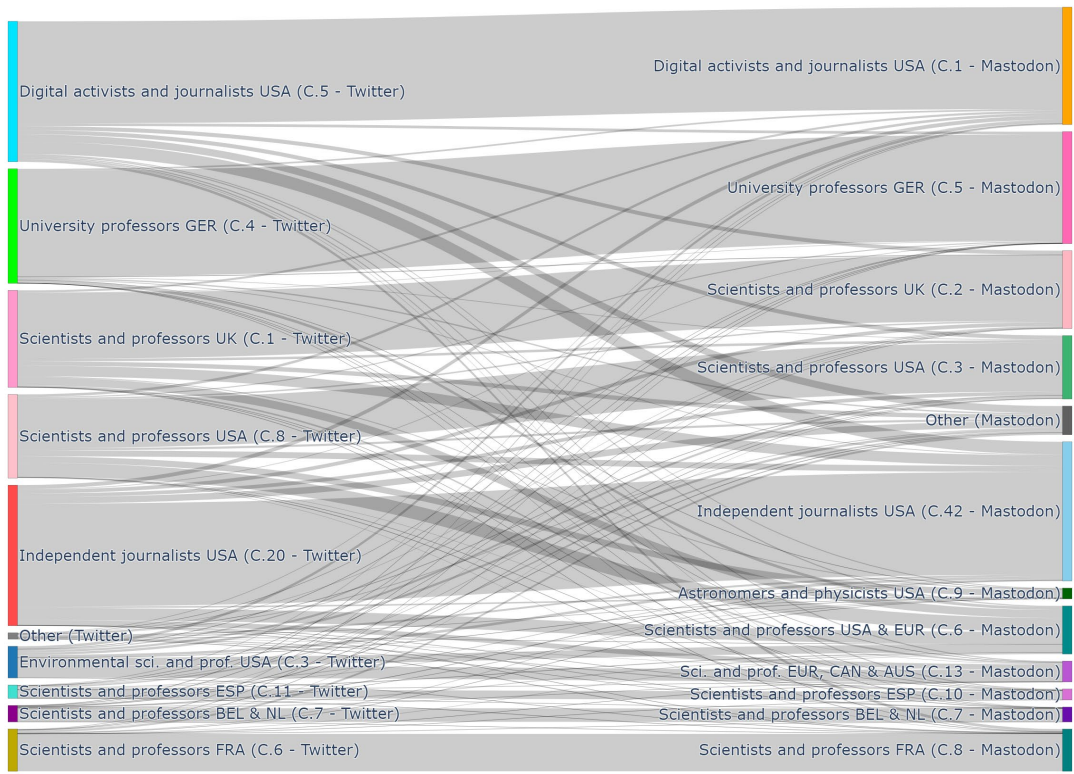

Source: Author's own work.

Table 2. The main community of origin (Twitter) per destination community (Mastodon) and percentage of the total.

| Community in Twitter (Origin) | Main Community(es) in Mastodon (Destination) | Percentage |
|---|---|---|
| Independent journalists USA (C.20) | Independent journalists USA (C.40) | 73.82% |
| Digital activists and journalists USA (C.5) | Digital activists and journalists USA (C.1) | 86.85% |
| University professors GER (C.4) | University professors GER (C.5) | 94.83% |
| Scientists and professors UK (C.1) | Scientists and professors UK (C.2) | 84.50% |
| Scientists and professors USA (C.8) | Scientists and professors USA (C.3) | 76.78% |
| | Scientists and professors USA and EUR (C.6) | 28.98% |
| | Astronomers and physicists USA (C.9) | 58.15% |
| Scientists and professors FRA (C.6) | Scientists and professors FRA (C.8) | 88.50% |
| Environmental sci. and prof. USA (C.3) | Sci. and prof. EUR, CAN, and AUS (C.13) | 57.26% |
| Scientists and professors BEL and NL (C.7) | Scientists and professors BEL and NL (C.7) | 90.97% |
| Scientists and professors ESP (C.11) | Scientists and professors ESP (C.10) | 76.10% |

Source: Author's own work.

Beyond the fact that the transfers or migrations from Twitter to Mastodon seem to happen in blocks, meaning that entire communities move and replicate their relationships as much as they can—not without losing a certain amount of relational capital along the way—it is striking that all the communities identified are academics, journalists, and activists from the US and Europe. Therefore, the users who led the campaign to leave came mostly from the intellectual and journalistic fields rather than from other areas, like politics or business.

*4.4. More Than a Migration, It Is About Straddling Between Both Platforms*

Finally, in order to test whether the concept of "migration" is a metaphor that correctly captures the phenomenon under analysis—after all, migration implies moving and leaving one place to relocate to another—we have analysed what percentage of authors in each community have tweeted before and after migration based on a sample of about 2000 users.

The data suggest, on the one hand, that on average, 98.49% of users posted content on Twitter during the 30 days prior to their registration in Mastodon, and 99.14% did so during the 30 days after. This implies a nominal increase of 0.91%, which makes it very difficult to refer to the phenomenon we are analysing as a migration. The number of users who stopped tweeting after registering in Mastodon is negligible, and in fact, some of them tweeted during the days after registering in Mastodon, while they did not do so during the previous days.

On the other hand, if we look at the intensity of publications, the data suggest that on average, there has been a modest decrease of 5.38% in tweet volume (from an average of 477.74 tweets per user per month to 452.02). However, the crucial finding supporting the 'straddling' concept is the persistence of substantial activity on Twitter for nearly all users in the sample (99.14% active post-registration). Even in the communities with the largest decreases, the average tweet volume remained remarkably high (e.g., over 590 tweets/month for UK scientists/professors). This sustained high engagement on the original platform, despite registering and presumably starting activity on Mastodon, strongly indicates the maintenance of active presences on both platforms, rather than a complete migration involving the abandonment of Twitter.

During the analysed period, none of the user communities included in the sample experienced a drastic reduction in the number of active users in the month following their transition to Mastodon. This indicates that Twitter users continued to participate regularly on this platform, which they began to use alongside the new one. The analysis of user activity in the sample also reinforces the idea that based on the volume of messages they post—averaging between 5 and 10 messages per day—these are very active communities of Twitter users.

## 5. Discussion and conclusions

The analysis of the migration of Twitter users to Mastodon leads to the conclusion, first and foremost, that this was a reactive movement led by a group of highly active users, mainly from the scientific, academic, and journalistic fields (Erz 2022). The correlation between the discussions and mentions on Twitter about the purchase of the social media by Elon Musk and the registration of users on Mastodon is clear. Regarding the motivations, it was driven by disagreement with the sale or with the new owner's policies, the perception that Twitter's principles were at risk, and that these could be better reflected in Mastodon's decentralised and collaborative structure (Jeong et al. 2024b).

This leadership exemplifies that subjective norm acts as a strong pull factor, shaping users' decisions not only through personal criteria but also through social pressures to maintain connections and follow peers' choices (Sun et al. 2017). Users with high activity, large social networks, and strong external visibility—so-called high net-worth individuals (HNWI)—play a leading role in social media migration, often setting trends and serving as early indicators of broader user abandonment (Kumar et al. 2011). In this case, the users who spearheaded this shift, clearly identifying themselves on Twitter through messages showing their move to Mastodon, formed a highly active community, with a proportion of followers and activity rates well above average. Other existing analyses indicate that migration occurred to a greater extent in communities with a strong collective identity and a high level of information exchange (La Cava et al. 2023). The content analysis conducted by He et al. (2023) of the posts on both platforms reaffirms this reactive and protest-driven component, as while users continued posting on a wide range of topics on Twitter, the content on Mastodon was more focused on discussions about the Fediverse and the migration from Twitter itself.

Secondly, the results also show that migrations can be very rapid in the digital environment, that virtual communities can shift rapidly from one platform to another, and that high volatility—particularly in terms of activity—is one of their defining characteristics.

Despite the ability of this group of users to move to the new platform with part of their community, the study shows that these users appear to have lost a relevant portion of their relational capital. The structural changes observed in the Mastodon networks suggest a more fragmented network structure compared to the users' original Twitter networks. While Mastodon's decentralised architecture offers benefits like user control and instance diversity, it may also complicate the direct replication of tightly knit communities from centralised platforms. This structural difference appears to contribute to the observed fragmentation, potentially impacting users' sense of community cohesion and increasing the challenge of rebuilding their established social capital (Zulli et al. 2020).

In the rapid search for a viable alternative to Twitter, Mastodon emerged as a natural option. However, there is no innovation without social appropriation, and the technical differences between both platforms, along with the difficulty in transferring accumulated capital, have hindered a true migration between two platforms that inherently had disparate technical characteristics and purposes. The transfer of highly active users inevitably involves a loss, which corroborates that technical properties also have social consequences (Laser et al. 2023) and, as users themselves point out, an accumulated social capital that is hard to leave behind (Kupferschmidt 2022).

Mastodon replicates, without commercial intent, the formal characteristics of mainstream platforms like Twitter or Reddit, but due to its features, it tends to create niche communities driven by technology enthusiasts, activists, students, and academics (Boyle et al. 2021). On one hand, Gehl and Zulli define as "technoelitism" the absence of a centralised registration point, and they argue that the need to make initial choices (such as selecting an instance) to register frustrates the expectations of general users, particularly those coming from more user-friendly commercial platforms (Gehl and Zulli 2022). This technological gap explains why Mastodon remains a niche for open-source software enthusiasts and people with greater technical ease, which is reflected in the abundance of topics related to technology and programming (Zignani et al. 2019).

In line with Bansal et al. (2005), while push factors such as dissatisfaction with Twitter's new ownership and moderation policies and pull factors such as Mastodon's perceived alignment with open-source values were present, mooring factors like users' accumulated social capital and technological barriers appear to have prevailed, helping to explain why the migration remained incomplete. In digital environments, staying behaviour is not always the result of satisfaction or rational loyalty; it can also arise from psychological, emotional, and practical barriers that keep users anchored to their current service (Sun et al. 2017).

For all the above reasons, the migration from Twitter to Mastodon can be characterised as having one foot in each camp or straddling both platforms. As Jeong et al. (2024a) point out, rather than full transitions, these migrations often result in parallel presences across platforms, shaped by professional needs, social dynamics, and perceived platform affordances. If we understand migration as leaving the space previously occupied, the users analysed took the first step towards the change, but without fully leaving their original platform behind. In other words, they began to engage significantly on Mastodon without abandoning their activity on Twitter. Along with the loss of relational capital, a more niche audience and contents in Mastodon and the uncertainty regarding its development may be considered as factors for this unfinished migration.

Several limitations should be acknowledged when interpreting these findings. It is important to recognise the specific nature of our sample. By focusing on Twitter users who publicly advertised their Mastodon accounts in their profiles, our study captures a self-selected group that was highly motivated and visible during the migration wave. The prevalence of academics, scientists, and journalists in this group, while aligning with observations from other studies (Braun 2023; Kupferschmidt 2022; Bittermann et al. 2023), means our findings primarily reflect the behaviour and network dynamics of these specific professional communities. The results may not be directly generalisable to the broader population of Twitter users or all Mastodon adopters. Furthermore, the reliance on web scraping for Mastodon data, necessitated by API constraints for accessing federated follow graphs comprehensively at the time, means our analysis is limited to publicly available data and excludes users on private instances or those with protected accounts. This introduces a potential selection bias towards users on more open instances and may affect the generalisability of the network structure findings. These factors restrict the generalisability of our results beyond the specific, highly engaged communities studied, primarily within academic and journalistic fields active on public Mastodon instances.

Additionally, the study's temporal scope presents limitations. The analysis of Twitter activity changes was based on a one-month window before and after Mastodon registration. While this timeframe is suitable for capturing the immediate reaction and testing the hypothesis of platform abandonment versus straddling, it is relatively short. It does not account for longer-term adaptation, potential learning curves associated with the new platform, or the gradual migration of users' social ties, all of which could influence activity levels over extended periods. Moreover, the dataset concludes in late 2022, capturing the peak migration period following Musk's acquisition. This study provides a snapshot of that critical moment but cannot address the evolution of these users' platform usage patterns beyond that point, which would allow for a more comprehensive analysis of the diffusion of innovation process (Rogers 2003). Whether the observed 'straddling' represents a stable, long-term strategy for these users or merely a transitional phase before eventually consolidating onto one platform (or abandoning both) remains an open question. Future longitudinal research is needed to track the evolution of platform usage and network structures over a longer period.

Despite the strong sensitivity and willingness for change, it is not enough for the elites of highly active users in a field to express their desire to migrate without the backing of the general public. The technical characteristics of the platforms affect their social use, and the accumulated capital acts as a counterbalance to the desire for change. However, the case of the failed migration from Twitter to Mastodon opens the possibility for potential rapid adoption in the future under similar circumstances if technological access barriers are reduced, enabling easier adoption by non-specialised users.